\documentclass[prl,aps,twocolumn,showpacs,groupedaddress,amsmath,amssymb]{revtex4}

\usepackage{graphicx}
\usepackage{dcolumn}
\usepackage{bm}
\usepackage{subfigure}
\usepackage{color}
\usepackage{verbatim}

\begin{document}

\title{
Spin chain in magnetic field:
limitations of the large-$N$ mean-field theory}

\author{Krzysztof Wohlfeld$^{1, 2\footnote{email address: krzysztof.wohlfeld@fuw.edu.pl}}$, Cheng-Chien Chen$^{3}$, 
Michel van Veenendaal$^{3,4}$, Thomas P. Devereaux$^{1}$
}
\affiliation{$^{1}$Stanford Institute for Materials and Energy Sciences, SLAC National Laboratory and Stanford University, 
Menlo Park, California 94025, USA}
\affiliation{$^{2}$Institute of Theoretical Physics, Faculty of Physics, University of Warsaw, Pasteura 5, PL-02093 Warsaw, Poland}
\affiliation{$^{3}$Advanced Photon Source, Argonne National Laboratory, Argonne, Illinois 60439, USA}
\affiliation{$^{4}$Department of Physics, Northern Illinois University, De Kalb, Illinois 60115, USA}
\date{\today}

\begin{abstract}
Motivated by the recent success in describing the spin and orbital spectrum of a spin-orbital chain
using a large-$N$ mean-field approximation~\cite{Chen2014}, we apply the same formalism to the case of a spin chain
in the external magnetic field. It occurs that in this case, which corresponds to $N=2$ in the approximation,  
the large-$N$ mean-field theory cannot qualitatively reproduce the spin excitation spectra at high magnetic
fields, which polarize more than 50\% of the spins in the magnetic ground state.
This, rather counterintuitively, shows that the physics of a spin chain can under some 
circumstances be regarded as more complex than the physics of a spin-orbital chain.
\end{abstract}


\maketitle

{\it Introduction}
Recently a number of  studies discussed the collective excitations 
in a spin-orbital chain~\cite{Wohlfeld2011, You2012, Schlappa2012, Bisogni2013, Chen2014}.
Most of them concentrated around a novel phenomenon
called spin-orbital separation which is present when a very strong external crystal field fully polarizes the orbital sector of the 
ground state~\cite{Wohlfeld2011, Schlappa2012, Bisogni2013}.
Although this phenomenon seemed to be completely at odds with the physics
present in an SU(4)-symmetric spin-orbital chain (i.e. without external
crystal field)~\cite{Zhang1999}, a very recent paper discusses how to unify these
two seemingly different limits~\cite{Chen2014}. It occurs that a large-$N$ mean-field
theory~\cite{Baskaran1987, Affleck1988} surprisingly well describes
the spin and orbital spectra for {\it any} value of the crystal field
and thus explains the striking evolution of the spin and orbital spectra
with increasing external crystal field~\cite{Chen2014}.

As a result of this recent success of the large-$N$ mean-field
theory the following question arises:
could such a theory be equally successful in explaining the behavior of collective
spin excitations in a spin chain that is subject to external magnetic field?
While this might look like as an old problem, 
which should have been solved long time ago, to the best of our knowledge,
there exists no precise answer to this question in the literature.

\begin{figure*}[t!]
\begin{minipage}{\textwidth}
\includegraphics[width=0.98\textwidth]{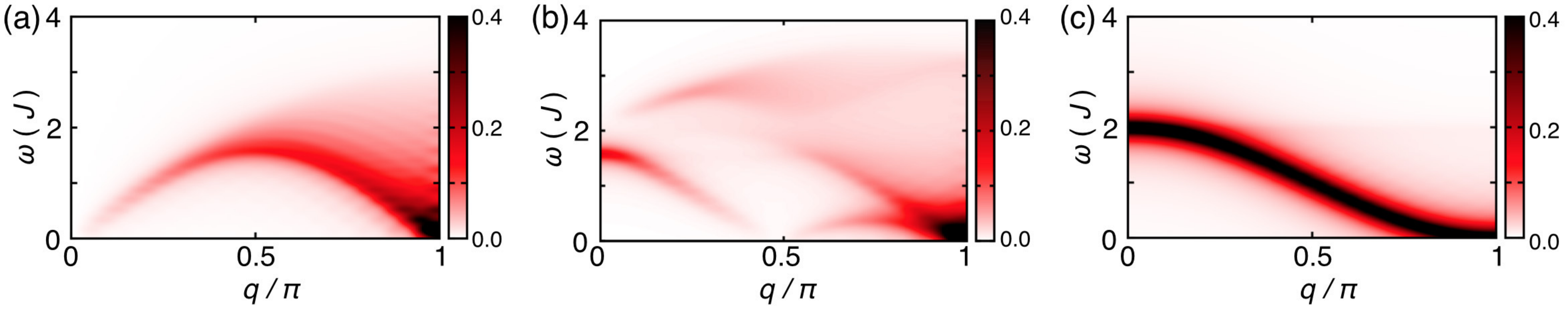}
\end{minipage}
\caption{{\it Numerical results} 
Spin dynamical structure factors computed by CPT+ED for a spin chain under a magnetic field $H_z$:
(a) $H_z=0$ with no spin polarization ($\sum_i S^z_i/L =0$);
(b) $H_z\sim 0.79 H^{cr}_z$ with half polarized spins ($\sum_i S^z_i/L =1/4$);
(c) $H_z =H^{cr}_z = 2.0J$ with fully polarized spins ($\sum_i S^z_i/L =1/2$).
The ED spectra (broadened with a 0.25$J$ Lorentzian) are computed on an $L=24$ site lattice.
For $H_z < H^{cr}_z$ [panels (a) and (b)], the spin spectra show fractionalized excitations with broad energy continua.
When $H_z\geq H^{cr}_z$, the (ferromagnetic) ground state is fully spin-polarized; the excitations are no longer fractional, and the spectrum exhibits only a sharp, single-magnon mode [panel (c)].
}
\label{fig:spinCPT}
\end{figure*}

{\it Definition of the problem}
Let us now be more specific. First, we define the following Hamiltonian which describes the problem of a spin chain
subject to the external magnetic field $H_z$ 
\begin{align} \label{eq:spin}
\mathcal{H} =& J \sum_{\langle ij \rangle} \Big( {\bf S}_i {\bf S}_j + \frac{1}{4} \Big) + H_z \sum_i S^z_i.
\end{align}
Here $J$ is the energy scale of the superexchange interactions between SU(2)-invariant spin $S=1/2$ operators (${\bf S}$),
$\langle ij \rangle$ represents a nearest-neighbor spin pair, and $H_z$ is the magnetic field strength.

Second, we  define the transverse dynamical spin structure factor which is a good
proxy for probing the nature of the collective spin excitations:
\begin{align} \label{eq:spindsf}
S(q, \omega)=\frac{1}{\pi} \lim_{\eta \rightarrow 0} 
\Im \langle \psi | S^x_{q} 
\frac{1}{\omega + E_{\psi}  - \mathcal{{H}} -
i \eta } S^x_q | \psi \rangle.
\end{align}
Here $| \psi \rangle$ is the ground state of $\mathcal{{H}}$ with energy $E_{\psi}$,
$S^x_q \equiv \sum_j e^{iqj} S^x_j /\sqrt{L}$ is the Fourier transform of the local spin operator, 
and $L$ is the number of lattice sites.

In what follows we calculate the dynamical spin structure factor in Eq. (\ref{eq:spindsf}) using
two distinct methods: (i) the numerically exact combined cluster perturbation theory (CPT)
and exact diagonalization (ED) method, and (ii) the approximate 
analytical large-$N$ mean-field theory. 
We compare our analytical results with the numerical calculations,
which have already been extensively discussed in the literature
~\cite{Mueller1981, Dender1997, Brenig2009, Kohno2009, Mourigal2013}.

{\it Numerical results}
The numerical method employed in the current study, CPT+ED, is a quantum cluster 
approach~\cite{Senechal2000, Maier2005} which complements the finite-size ED simulations
and therefore allows for a better visualization of the fine spectral details.
The spin dynamical structure factor, Eq. (\ref{eq:spindsf}), calculated with this method
is shown in Fig.~\ref{fig:spinCPT} for three different values of the external magnetic
field $H_z$. When $H_z=0$, the ground state has short range antiferromagnetic order
and the spectrum is well-known~\cite{Mueller1981, Dender1997, Brenig2009, Mourigal2013}: it is mostly
spanned by a two-spinon continuum and has zero modes at $q=0$ and $q=\pi$, cf. Fig.~\ref{fig:spinCPT}(a).
For a finite value of $H_z$ the ferromagnetic domains start to appear in the ground state and the positions
of the zero modes shift. For instance when $H_z \sim 1.58J$ half of the spins in the chain are polarized 
and the spectrum has zero modes at $q=\pi / 2$ and $q=\pi$, cf. Fig.~\ref{fig:spinCPT}(b). Nevertheless,
the spectrum is still spanned by a continuum of fractional excitations. This, however, stays in contrast with the fully polarized 
(i.e. ferromagnetic) ground state which occurs at $H_z = 2J \equiv H_z^{cr}$. The spin dynamical
structure factor is then no longer spanned by a two-spinon continuum, cf. Fig.~\ref{fig:spinCPT}(c), and instead
a single (magnon) branch arises. Further increasing $H_z$ above $H_z^{cr}$  leads to a gapped magnon branch.

\begin{figure}[t!]
\includegraphics[width=0.9\columnwidth]{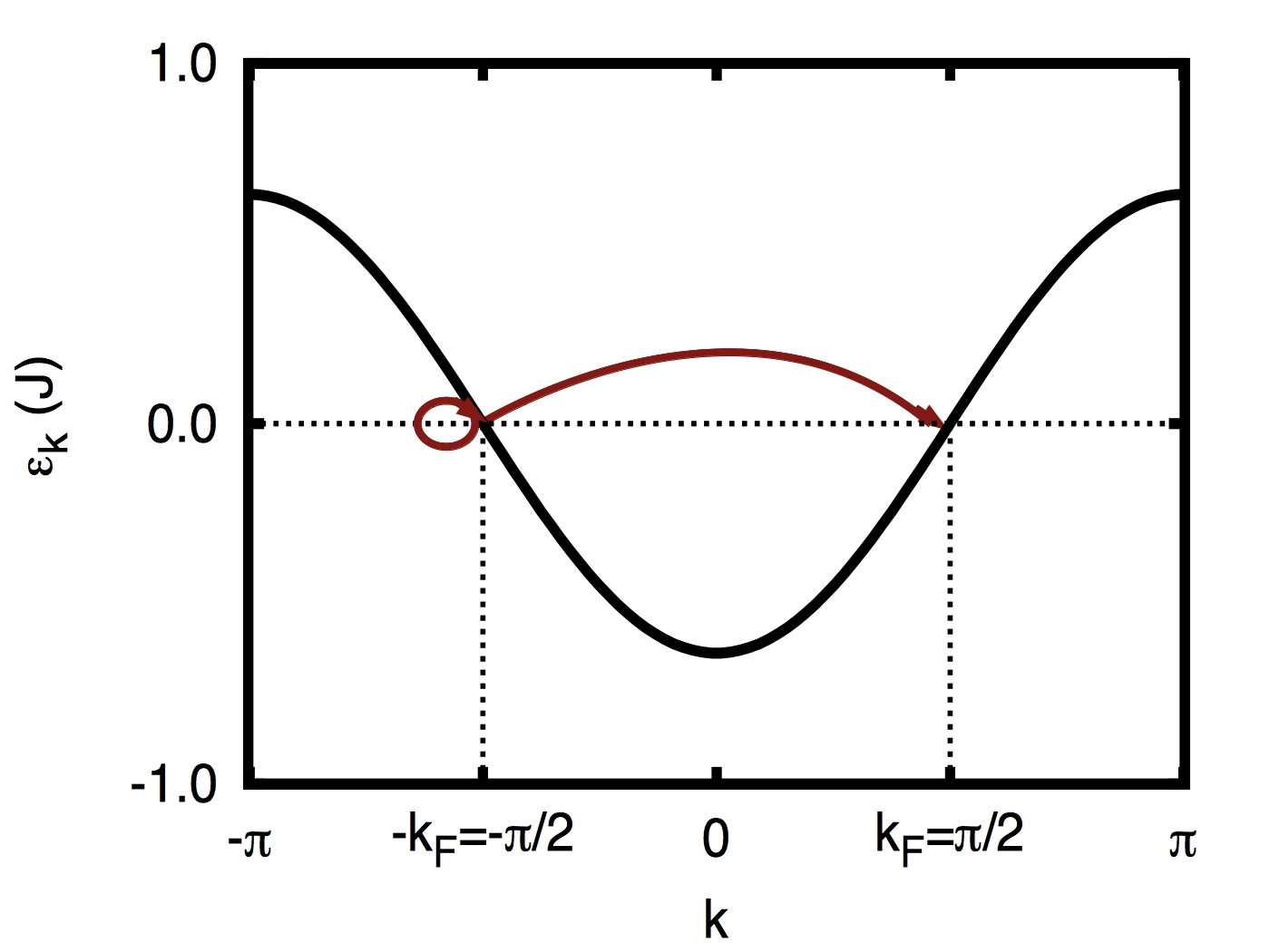} \\
\includegraphics[width=0.9\columnwidth]{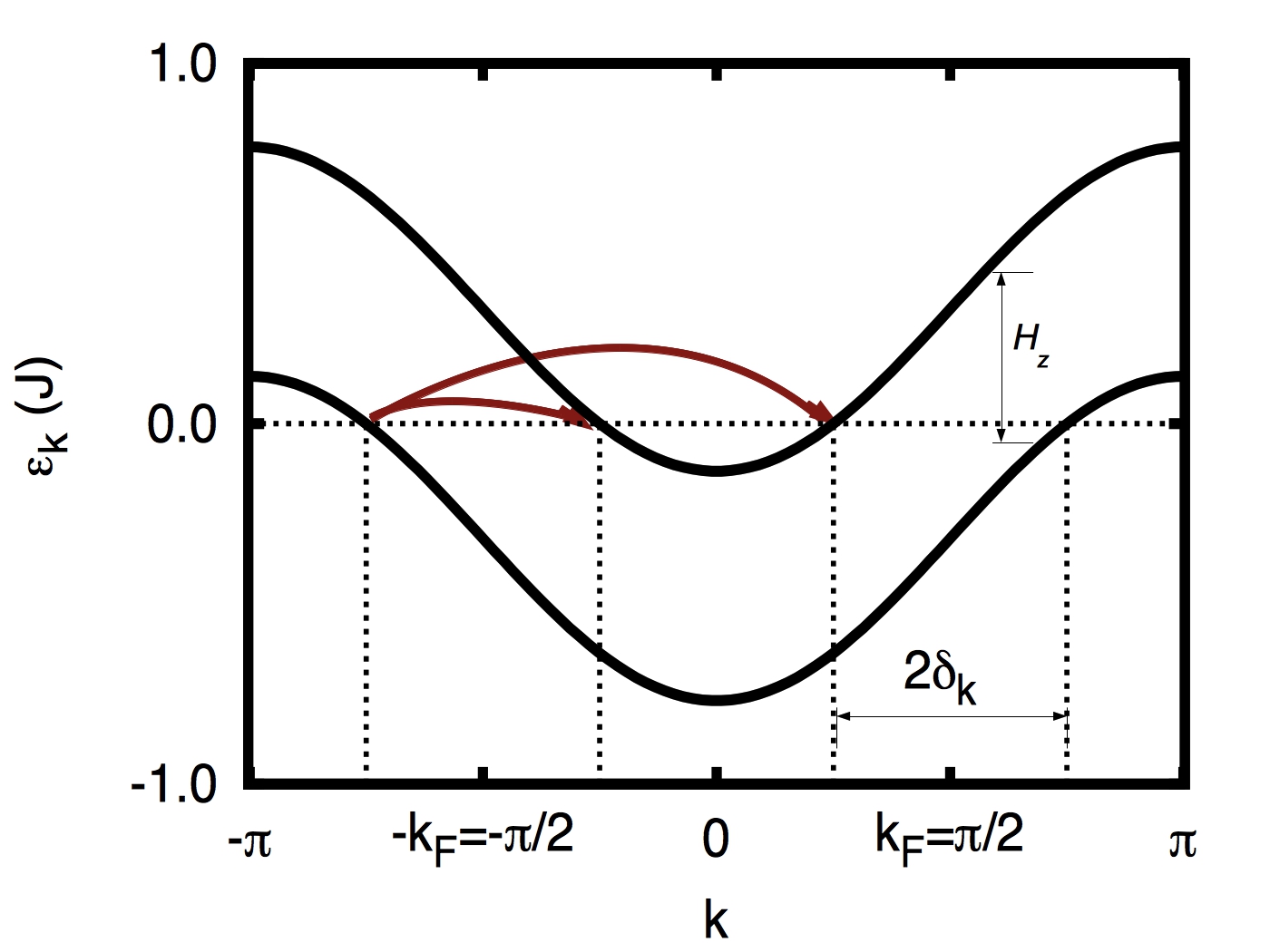}
\caption{{\it Large-$N$ mean-field theory} Evolution of the mean-field fermionic bands as a function of the magnetic field at $H_z=0$ (top panel) and $H_z= 2J / \pi$ (bottom panel).
The collective spin excitations in the mean-field picture correspond to `particle-hole' excitations of the fermions across the 
Fermi surface (denoted by the dotted horizontal lines).
The energies of the up-spin and  down-spin fermionic bands are separated by $H_z$,
and the allowed `particle-hole' excitations change with the magnetic field accordingly.
The thick arrows point to the allowed zero-energy spin excitations.
}
\label{fig:spinbands}
\end{figure}

\begin{figure*}[t!]
\begin{minipage}{\textwidth}
\includegraphics[width=0.98\textwidth]{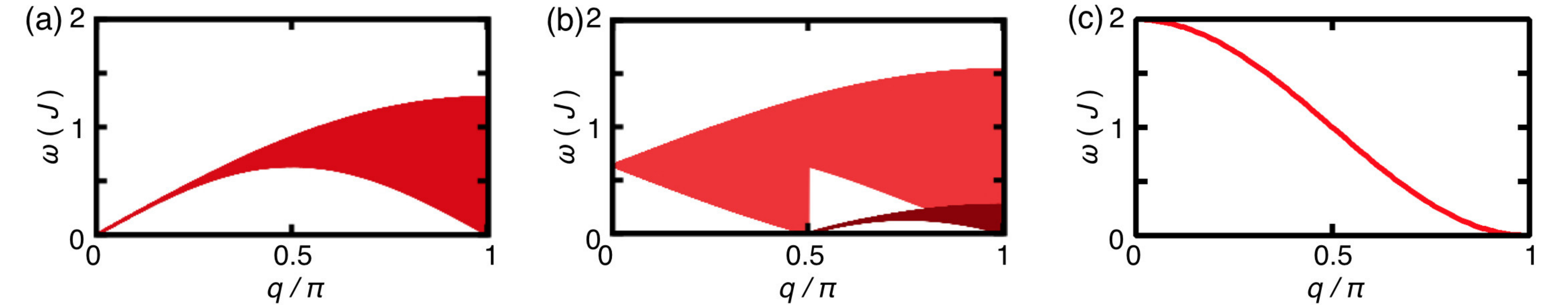}
\end{minipage}
\caption{{\it Analytical results}
Compact support of the spin spectra for a spin chain in a magnetic field $H_z$ computed by the large-$N$ mean-field theory [(a)-(b)]: 
(a) When $H_z =0$, the ground state exhibits (short-range) antiferromagnetic correlations without any spin polarization.
(b) When $H_z =  2 J / \pi $ in the mean-field calculation,  i.e. when 50\% of the spins are polarized in the ground state;
the lighter (darker) part in the spin spectrum refers to spin-flip excitations created by spin raising (lowering) operators.
(c) The exact spin wave dispersion $\omega_q=J(1+\cos q)$ calculated by linear spin wave theory at $H_z=H^{cr}_z$;
the large-$N$ mean-field theory is not valid in this regime.
}
\label{fig:mfspinspectra}
\end{figure*}

{\it Large-$N$ mean-field theory}
Following similar steps as described in detail in Refs.~\cite{Kenzelmann2005, Chen2014}
we first map the spin model described by Eq. (\ref{eq:spin}) onto a fermionic model~\cite{Kenzelmann2005}.
By performing a large-$N$ mean-field decoupling and solving the self-consistent mean-field equations,  we obtain the following Hamiltonian
 \begin{align}\label{eq:hfermionspin}
\mathcal{H}^{\rm MF} =
& \sum_{k }
 \big(  \varepsilon_{k \uparrow} f^\dag_{k \uparrow}  f_{k \uparrow} +  \varepsilon_{k \downarrow}
 f^\dag_{k \downarrow}  f_{k \downarrow} \big),
 \end{align} 
where the fermionic bands are $ \varepsilon_{k \uparrow / \downarrow} = -  2 J   \cos (\delta_k) \cos k / \pi  \mp J \sin(2 \delta_k) / \pi$ (cf. Fig.~\ref{fig:spinbands}) 
with $\delta_k = \arcsin [H_z \pi / (2J)] / 2 $, subject to the constraint $ \sum_{\sigma} f^\dag_{i \sigma}  f_{i \sigma} = 1$. 

We note, first, that in this approach $H_z \le 2J / \pi$,
which means $ \delta_k \le \pi / 4$ and a maximum Fermi momentum $\delta_k + k_F = 3 \pi /4$ under an applied magnetic field 
(since $k_F = \pi /2 $ for the fermionic mean-field theory at $H_z=0$~\cite{Kenzelmann2005}).
Second, above $H_z > 2 J / \pi$ the self-consistent mean-field solution effectively breaks down.

Next, since the maximum available Fermi momentum, $3 \pi /4$, corresponds to just half of the spins being polarized in the ground state, this is the maximum
value of polarization available in this approach. 
As this means that we can never fully fill one of the spin bands, i.e. reach the $\delta_k + k_F = \pi$ momentum, 
the fermionic mean-field theory is not able to describe the fully spin-polarized ground state.

In this fermionic mean-field picture, the compact support of the spin spectrum can be calculated by $\bar{S}(q, \omega) =  \sum_{k \in FS,  q+k \notin FS} \delta (\omega - \varepsilon_{q+k, \uparrow} + \varepsilon_{k \downarrow}) + \sum_{k \in FS,  q+k \notin FS} \delta (\omega - \varepsilon_{q+k, \downarrow} + \varepsilon_{k \uparrow})$, 
cf.~Fig.~\ref{fig:spinbands}.
The evolution of the spin spectrum as a function of $H_z$ is shown in Fig.~\ref{fig:mfspinspectra}(a)-(b).
The shift of the zero-energy modes with the magnetic field in the spin spectrum follows from the splitting of the fermionic bands in the 
magnetic field, cf.~Fig.~\ref{fig:spinbands}.

{\it Conclusions}
Let us first emphasize that the fermionic large-$N$ mean-field approximation is valid only when $H_z \leq 2 J / \pi$ (i.e. when 
the ground state is less than half-polarized).
In this case there exists a qualitative agreement between the analytically calculated compact support and the numerically obtained spin spectrum. 
Nevertheless, the quantitative differences between the mean-field and the CPT+ED results are substantial (the bandwidth of the spin excitations
is ca. 2.5 smaller in the analytical approach), 
making the fermionic large-$N$ mean-field description less appropriate compared to that in the spin-orbital model~\cite{Chen2014}.

A different situation exists when  $H_z > 2 J / \pi$  (i.e. when the ground state is more than half-polarized). 
Here the mean-field approximation breaks down. This means that  {\it long} before the critical field is reached (which fully polarizes the ground state)
there exists no {\it simple} analytical approach that can be used to calculate the spin spectrum (for more complex approaches, cf. Ref.~\cite{Mueller1981, Karbach2002}). 
It is then only when $H_z \geq H^{cr}_z$
that another simple analytical approach becomes valid -- the well-known linear spin-wave approximation which works well for spin ground states
with long range magnetic order~\cite{Brenig2009, Auerbach1994}. 
The analytically evaluated spin spectrum, $ S(q, \omega) = \delta (\omega - \omega_q)$ with $\omega_q = J (1+ \cos q )$ being the 
magnon dispersion when $H_z = H^{cr}_z$ [cf. Fig.~\ref{fig:mfspinspectra}(c)], is then equal to the numerically calculated spin dynamical structure factor [cf. Fig.~\ref{fig:spinCPT}(c)]. 

{\it Acknowledgements}
K. W. and T. P. D. acknowledge support from the DOE-BES Division of Materials Sciences and Engineering (DMSE) under Contract No. DE-AC02-76SF00515 (Stanford/SIMES). 
K. W. acknowledges support from the Polish National Science Center (NCN) under Project No. 2012/04/A/ST3/00331.
C.C.C. is supported by the Aneesur Rahman Postdoctoral Fellowship at Argonne National Laboratory (ANL), operated by the U.S. DOE Contract No. DE-AC02-06CH11357.
M.v.V is supported by the DOE Office of BES Award No. DE-FG02-03ER46097 and the NIU Institute for Nanoscience, Engineering and Technology.
This work utilized computational resources at NERSC, supported by the U.S. DOE Contract No. DE-AC02-05CH11231.


\begin{thebibliography}{10}

\bibitem{Chen2014}
C.-C. {Chen}, M. {van Veenendaal}, T.~P. {Devereaux}, and K. {Wohlfeld}, Phys. Rev. B {\bf 91},  165102 (2015).

\bibitem{Wohlfeld2011}
K. Wohlfeld, M. Daghofer, S. Nishimoto, G. Khaliullin, and J. ~v.~d. Brink,
Phys. Rev. Lett. {\bf 107},  147201 (2011).

\bibitem{Schlappa2012}
J. Schlappa, K. Wohlfeld, K.~J. Zhou, M. Mourigal, M.~W. Haverkort, V.~N.
  Strocov, L. Hozoi, C. Monney, S. Nishimoto, S. Singh, A. Revcolevschi, J.
  Caux, L. Patthey, H.~M. R{\o}nnow, J.~v.~d. Brink, and T. Schmitt, Nature {\bf
  485},  82  (2012).

\bibitem{Bisogni2013}
V. {Bisogni}, K. {Wohlfeld}, S. {Nishimoto}, C. {Monney}, J. {Trinckauf}, K.
  {Zhou}, R. {Kraus}, K. {Koepernik}, C. {Sekar}, V. {Strocov}, B. {Buechner},
  T. {Schmitt}, J. {van den Brink}, and J. {Geck}, Phys. Rev. Lett. {\bf 114}, 096402 (2015).
  
 \bibitem{You2012}
W.-L. You, A. M. Ole\'s, and P. Horsch, Phys. Rev. B {\bf 86}, 094412 (2012);
W. Brzezicki, J. Dziarmaga, and A. M. Ole\'s, Phys. Rev. Lett. {\bf 112}, 117204 (2014).

\bibitem{Zhang1999}
Y.-Q. Li, M. Ma, D.-N. Shi, and F.-C. Zhang, Phys. Rev. B {\bf 60},  12781
  (1999).

\bibitem{Baskaran1987}
G. Baskaran, Z. Zou, and P. Anderson, Solid State Communications {\bf 63},  973
    (1987).

\bibitem{Affleck1988}
I. Affleck and J.~B. Marston, Phys. Rev. B {\bf 37},  3774  (1988).

\bibitem{Mueller1981}
G. M\"uller, H. Thomas, H. Beck, and J.~C. Bonner, Phys. Rev. B {\bf 24},  1429
   (1981).

\bibitem{Dender1997}
D.~C. Dender, P.~R. Hammar, D.~H. Reich, C. Broholm, and G. Aeppli, Phys. Rev.
  Lett. {\bf 79},  1750  (1997).

\bibitem{Brenig2009}
S. Grossjohann and W. Brenig, Phys. Rev. B {\bf 79},  094409  (2009).

\bibitem{Kohno2009}
M. Kohno, Phys. Rev. Lett. {\bf 102},  037203  (2009).

\bibitem{Mourigal2013}
M. {Mourigal}, M. {Enderle}, A. {Kl{\"o}pperpieper}, J.-S. {Caux}, A.
  {Stunault}, and H.~M. {R{\o}nnow}, Nat. Phys. {\bf 9},  435  (2013).

\bibitem{Senechal2000}
D. S\'en\'echal, D. Perez, and M. Pioro-Ladri\`ere, Phys. Rev. Lett. {\bf 84},
  522  (2000).

\bibitem{Maier2005}
T. Maier, M. Jarrell, T. Pruschke, and M.~H. Hettler, Rev. Mod. Phys. {\bf 77},
   1027  (2005).

\bibitem{Kenzelmann2005}
M. Kenzelmann, C.~D. Batista, Y. Chen, C. Broholm, D.~H. Reich, S. Park, and Y.
  Qiu, Phys. Rev. B {\bf 71},  094411  (2005).

\bibitem{Karbach2002}
M. Karbach, D. Biegel, and G. M\"uller, Phys. Rev. B {\bf 66},  054405  (2002).

\bibitem{Auerbach1994}
A. Auerbach, {\em Interacting Electrons and Quantum Magnetism} (Springer, New
  York, 1994).

\end{thebibliography}

\end{document}